
%
\input phyzzx
%
%
\newcount\lemnumber   \lemnumber=0
\newcount\thnumber   \thnumber=0
\newcount\conumber   \conumber=0

\def\myeq{{\the\equanumber}}

\def\Lemma{\par\noindent\global\advance\lemnumber by 1
           {\bf Lemma\ (\chapterlabel\the\lemnumber)}}
\def\Corollary{\par\noindent\global\advance\conumber by 1
           {\bf Corollary\ (\chapterlabel\the\conumber)}}
\def\Theorem{\par\noindent\global\advance\thnumber by 1
           {\bf Theorem\ (\chapterlabel\the\thnumber)}}

%
%
\def\e{\adveq\eqno{\rm (\chapterlabel\the\equanumber)}}
\def\adveq{\global\advance\equanumber by 1}


%
%
\font\tensl=cmsl10
\font\tenss=cmssq8 scaled\magstep1
\outer\def\quote{
   \begingroup\bigskip\vfill
   \def\endquote{\endgroup\eject}
    \def\par{\ifhmode\/\endgraf\fi}\obeylines
    \tenrm \let\tt=\twelvett
    \baselineskip=10pt \interlinepenalty=1000
    \leftskip=0pt plus 60pc minus \parindent \parfillskip=0pt
     \let\rm=\tenss \let\sl=\tensl \everypar{\sl}}
\def\from#1(#2){\smallskip\noindent\rm--- #1\unskip\enspace(#2)\bigskip}

\def\CALT{\address{Division of Physics, Mathematics
and Astronomy\break
Mail Code 452--48\break
California Institute of Technology\break
Pasadena, CA 91125}}

\def\r#1{$\lb \rm#1 \rb$}

%
%
\def\rarrow{\rightarrow}

\def\semidirect{\mathrel{\raise0.04cm\hbox{${\scriptscriptstyle |\!}$
\hskip-0.175cm}\times}}


\def\ref#1{$^{#1}$}

\def\wiggle{\tilde}

\def\Res{\mathop{\rm Res}\limits}

\def\half{{1\over2}}
\def\lb{\lbrack}
\def\rb{\rbrack}

\def\diam{{\hbox{\hskip-0.02in
\raise-0.126in\hbox{$\displaystyle\bigvee$}\hskip-0.241in
\raise0.099in\hbox{ $\displaystyle{\bigwedge}$}}}}

\def\bw#1#2#3#4#5{{w\left(\matrix{#1&#2\cr#3&#4\cr}\bigg\vert #5\right)}}

\overfullrule=0pt
\date{March, 1994}
\titlepage
\title{Spectra of RSOS Soliton Theories}
\author{Doron Gepner\foot{On leave from the Weizmann Institute, Israel.}}
\CALT
\vskip15pt
\abstract
We study here the spectrum of soliton scattering theories based on interaction
round the face lattice models. We take for the admissibility condition the
fusion rules of each of the simple Lie algebras. It is found that the
mass spectrum
is given by that of the corresponding Toda theory, or, that the mass ratios
of the different kinks in the model are described by the Perron--Frobenius
vector of the Cartan matrix. The scalar part of the soliton amplitude is
shown to be identical with the minimal part of the corresponding Toda
amplitude.
\endpage
%
The factorization property of integrable field theories in two dimensions
\REF\ZamZam{A.A. Zamolodchikov and Al.A. Zamolodchikov, Annals of Physics 120
(1979) 253}
\r\ZamZam\
provided an excellent way to study their $S$ matrices. The constraints
arising from factorization are
\REF\Vega{H.J. de Vega and V.A. Fateev, Int. Journal of Mod. Physics A6 (1991)
3221}
in many cases enough to determine them. In ref. \r\Vega\ the $S$ matrices of
$SU(N)$
RSOS theory were determined using the Boltzmann weights of the corresponding
lattice
models. Our purpose here is to extend this to other Lie algebras, and to
determine
especially the mass spectra of the solitons, continuing the work
\REF\RSOS{D. Gepner, Phys. Lett. B 313 (1993) 45}
of ref. \r\RSOS.

As an ingredient, we use the Boltzmann weights found in ref.
\REF\found{D. Gepner, Caltech preprint CALT--68--1825, November (1992)}
\r\found. Most of which,
however, were not proven to satisfy the YBE and are only conjectured to do so.
Part of our aim is to see whether a consistent soliton theory arises, which is
a
highly non--trivial check on these Boltzmann weights.
\par
The following expression for the Boltzmann weights of the fusion interaction
round the face (IRF) lattice model was found in ref. \r\found. The model is
denoted
by IRF$({\cal O}, x,x)$, where $\cal O$ is some rational conformal field
theory,
(RCFT), and $x$ is a primary field in it. The solution to the Young Baxter
equation (YBE) associated to this lattice model is
$$X_i(u)=\sum_{a=0}^{n-1} f_a(u) P^a_i,\e$$
where
$$f_a(u)=\prod_{r=0}^{n-a-2} \sin(\zeta_r+u) \prod_{r=n-a-1}^{n-2} \sin(\zeta_r
-u),\e$$
where
$$\zeta_a={\pi\over2} [\Delta(\phi_{a+1})-\Delta(\phi_a)],\e$$
where $\phi_i$ are the fields appearing in the operator product of $x$ with
respect
to itself, arranged in increasing dimensions,
$$x \cdot x=\sum_{i=0}^{n-1} \phi_i,\e$$
and $\Delta(\phi)$ is the conformal dimension of $\phi$.
The $P^a_i$ are the eigenvectors of the braiding matrices of the RCFT, braided
with the field $x$, and projected to the $a$ intermediate field in the
$t$ channel (see ref. \r\found, for detail).
The $X_i(u)$ so defined obey the Young Baxter equation
$$X_i(u) X_{i+1}(u+v) X_i(v)=X_{i+1}(v) X_{i}(u+v) X_{i+1}(u).\e$$
Actually, the fact that $X_i(u)$ so defined obeys the Young Baxter equation
eq. (5) has been proved only for $n=2$ \r\found, mainly,
and is conjectured that it is so in general.

Part of the motivation here
in exploring this solution is to examine whether this conjecture is
plausible. This we do by studying soliton theories based on the anzats eq. (1),
calculating their mass spectrum and comparing it to known systems.
In a nutshell, let $G$ stand for the WZW RCFT associated to the Lie group
$G$, and let $x$ be the fundamental representation. Then it is found here that
the soliton theory based
on $X_i(u)$ has a mass spectrum which is identical to the classical masses
of the $G$ Toda theory.

The soliton scattering amplitude associated to IRF$({\cal O},x,x)$ is given by
\r\found,
$$S^{ab}_{cd}(\theta) =F(\theta) \left( {S_{a,0} S_{c,0}\over S_{b,0} S_{d,0}}
\right)^{\theta/2} \bw b c a d {\lambda\theta}.\e$$
Here $S^{ab}_{cd}(\theta)$ is the scattering matrix of the kinks $K_{ab}+K
_{bc}\rarrow K_{a d}+K_{d c}$, at the relative rapidity $i\pi \theta$,
where $K_{ab}$ is the kink interpolating the $a$
and $b$ vacua. $S_{ab}$ is the torus modular matrix and
$\bw a b c d u$ is the Boltzmann weight associated to $X_i(u)$,
$$\bw a b c d u= \langle a c d | X_2(u) | a b d \rangle.\e$$
$F(\theta)$ obeys,
$$F(\theta)F(-\theta)=\prod_a \sin(\zeta_a-\lambda\theta)^{-1}
\sin(\zeta_a+\lambda\theta)^{-1}=\rho(\theta)\rho(-\theta),\e$$
along with
$$F(\theta)=F(1-\theta),\e$$
if $x$ is a real primary field, $\bar x=x$, or otherwise
$$F(1-\theta) F(1+\theta)=\prod_a \sin(\bar \zeta_a-\lambda\theta)^{-1} \sin(
\bar \zeta+\lambda\theta)^{-1}=\bar\rho(\theta)\bar\rho(-\theta),\e$$
where $\bar \zeta_a=\pi[\Delta(\bar\phi_{a+1})-\Delta(\bar\phi_a)]$,
and $\bar\phi_a$ are the primary fields appearing in the operator product
$$x\cdot \bar x=\sum_{a=0}^m \bar\phi_a,\e$$
arranged according to increasing dimensions. $\lambda=\bar\zeta_1$ is the
crossing parameter.
The function $F(\theta)$
is determined by eqs. (8,10) up to an ambiguity, called c.d.d. ambiguity, i.e.,
$\wiggle F(\theta)$ solves eqs. (8,10), if $F(\theta)$ does, $Q(\theta)$ is
any holomorphic function, and
$$\wiggle F(\theta)=F(\theta){Q(\theta) Q(1-\theta) \over Q(-\theta)
Q(1+\theta)}.\e$$
However, the assumption of minimality of the solution can be justified
in most cases: $F(\theta)$ is the solution of this equations with the
minimal number of poles and zeros on the physical regime $0\leq \theta < 1$.

The bound states of the two kinks $x$ appear as poles in the scattering
amplitude for the rapidities $0<\theta<1$. These, in turn are determined
by the function $F(\theta)$. We shall assume, owing to minimality that these
are exactly the poles of $\rho(\theta)$ or $\bar \rho(1-\theta)$. In
other words, $F(\theta)$ has poles at
$$\theta={\zeta_i\over\lambda},\qquad{\rm or\ } \theta_i=
1-{\bar\zeta_i\over\lambda}.\e$$

We can now compute the bound states spectrum, under the above assumptions,
for any RCFT and any primary field $x$. Importantly, this does not require
the explicit knowledge of the Boltzmann weights, eq. (1).
Our purpose is to check whether
this leads to a sensible soliton scattering theory. We shall see that
this is indeed so, by examining the different algebras, case by case.

{\it $A_n$ case.} Denote by $\Lambda_m$ the $m$'th fundamental weight of
the Lie algebra
$A_n$, $m=1,2,\ldots,n$. Consider the model IRF$(A_n,\Lambda_m,\Lambda_m)$.
We need to compute the parameters $\zeta_m$ and $\bar\zeta_m$. Thus,
consider the two tensor products,
$$[\Lambda_m] \times [\Lambda_m]=\sum_{r=0}^m [\Lambda_{2m-r}+\Lambda_r],\e$$
$$[\Lambda_m]\times[\bar\Lambda_m]=\sum_{r=0}^m [\Lambda_r+\bar\Lambda_r],\e$$
where we arranged the representations on the right hand side according to
increasing conformal dimensions. Using the dimension formula for WZW
primary fields,
$$\Delta={\lambda(\lambda+2\rho)\over k+g},\e$$
where $\lambda$ is the highest weight, $\rho$ is half the sum of positive
roots, $g$ is the dual Coxeter number and $k$ is the level, we find
$$\zeta_r={\pi\over2} (\Delta_{[\Lambda_{r+1}+\Lambda_{2m-r-1}]}-
\Delta_{[\Lambda_r+\Lambda_{2m-r}]})={\pi\over2} {2(m-r)\over k+n+1},\e$$
$$\bar \zeta_r={\pi\over2}(\Delta_{[\Lambda_{r+1}+\bar\Lambda_{r+1}]}-
\Delta_{[\Lambda_r+\bar\Lambda_r]})={\pi\over2} {n+1-2r\over k+n+1}.\e$$
The crossing parameter is $\lambda=\bar\zeta_1={\pi(n+1)\over 2(k+n+1)}$.
Thus, the poles in the physical regime of $F(\theta)$ are located at
$\zeta_r/\lambda$ and $1-\bar\zeta_r/\lambda$ and are given by,
$${2(r+1)\over n+1},\qquad r=0,1,\ldots,m-1,\e$$
and
$${2r\over n+1} \qquad {\rm for \ } r=1,2,\ldots,m-1.\e$$
Denote by
$$f_\alpha(\theta)={\sin[{\pi\over2}(\alpha+\theta)] \over
                   \sin[{\pi\over2} (\alpha-\theta)] }.\e$$
Then $F(\theta)$ can be written as,
$$F(\theta)=A_m \wiggle F(\theta),\e$$
where $\wiggle F(\theta)$ has no zeros or poles in the physical domain,
and is the unique such solution of eqs. (8,10) with this property. $A_m$ is
given by
$$A_m=f_{2/(n+1)}(\theta)^2 f_{4/(n+1)}(\theta)^2\ldots
f_{2(m-1)/(n+1)}(\theta)^2 f_{2m/(n+1)}(\theta).\e$$

Note, that $A_m$ is exactly the scattering amplitude for two $[m]$ particles
in the diagonal $A_n$ Koberle-Swieca
\REF\KS{R. Koberle and J.A. Swieca, Phys.Lett. 86B (1979) 209}\r\KS\
amplitude. This is essentially also
the scattering matrix of the affine $A_n$ Toda theory. Thus also the mass
spectrum of the particles are identical. The mass of the $\Lambda_m$ kink,
as well as the $[m]$ particle in the Toda theory are
$$M_m=\sin\left({\pi m\over n+1}\right), \qquad m=1,2,\ldots,n.\e$$
The unique simple pole in $A_m(\theta)$, at $\theta=2m/(n+1)$ corresponds
to the unique bound states in this channel, $[m]+[m]\rarrow[2m]$.
In general a pole at $0<\theta_p<1$ corresponds to a bound states mass of
$$M_B=2M\cos\left({\pi \theta_p \over 2}\right).\e$$
We, thus, conclude that up to the `$Z$ factor' $\wiggle F(\theta)$, the
scalar part of the $A_m$ kink theory is identical to the Koberle--Swieca
$A_n$ amplitude, with the $[m]$ particle identified with the $\Lambda_m$
kink.

Another example of an $A_n$ RSOS scattering theory is obtained by taking the
lattice model IRF$(A_n,x,x)$, where $x$ is the representation
with the highest weight $x=[m\Lambda_1]$. We can again calculate by the same
method the crossing parameters. The relevant tensor products are
$$[m\Lambda_1]\times [m\Lambda_1]=[m\Lambda_2]+[(m-1)\Lambda_2+2\Lambda_1]+
\ldots,\e$$
$$[m\Lambda_1]\times [m\bar\Lambda_1]=[0]+[\Lambda_1+\bar\Lambda_1]+
\ldots,\e$$
where we indicated only the representations that are relevant for the poles
in the physical sheet. We now find the parameters, by calculating the
conformal dimensions,
$$\lambda=\bar\zeta_0={\pi\over 2} {n+1\over k+n+1},\qquad\zeta_0=
{\pi \over 2} {2\over k+n+1}.\e$$
Thus, the unique pole on the physical sheet is at $\theta_p=2/(n+1)$.
It follows that again the amplitude can be written as
$$F(\theta) =A_1(\theta) \wiggle F(\theta),\e$$
where $A_1(\theta)$ has all the zeros and poles in the physical sheet,
and
$$A_1(\theta)=f_{2/(n+1)}(\theta),\e$$
Similarly on could consider the kink based on IRF$(A_n,x,x)$, where
$x=[m\Lambda_s]$ is the representation with the highest weight $m\Lambda_s$.
We find, in the same manner, the amplitude $A_s$. We conclude that
independently
of $m$, this scattering theory has the Koberle--Swieca mass spectrum,
eq. (24), with $m$ identified with $s$, and that the scalar amplitudes
are given precisely by the Koberle--Swieca scattering amplitudes. This implies
also that the interaction among the representations is identical to that
of the Koberle--Swieca.

An interesting question is which integrable theory is described by these
amplitudes. For $m=1$ it has been conjectured that this is the
RCFT $SU(n+1)_1 \times SU(n+1)_k\over SU(n+1)_{k+1}$, perturbed by the
field $\phi^{0,0}_{\rm ad}$ \r\Vega. However, we
see that for any $m$ we get the same masses, and consequently, the
same spins of the integrals of motion. Thus, it remains a mystery what is the
corresponding perturbed RCFT.

{\it $D_m$ case}. Let us turn now to the calculation of the poles for the $D_m$
scattering theory. Consider the model IRF$(D_m,s,s)$ where $s$ stands for the
spinor representation. We choose a basis in which the simple roots are
$\alpha_i=\epsilon_i-\epsilon_{i+1}$, $i=1,2,\ldots,m-1$,
$\alpha_m=\epsilon_i+\epsilon_{i+1}$, and $\epsilon_i$ form an orthonormal
set of unit vectors. The spinor weight is $\lambda=(\epsilon_1+\epsilon_2+
\ldots+\epsilon_m)/2$. The relevant tensor products are now, for even
$m$,
$$s\times s=\sum_{r=0}^{m/2}  [\epsilon_1+\epsilon_2+\ldots+\epsilon_{2r}],\e$$
and $s$ is a real representation.
For odd $m$ we find,
$$s\times s=\sum_{r=0}^{(m-1)/2} [\epsilon_1+\epsilon_2+\ldots+\epsilon_{2r-1}]
\e$$
$$s\times \bar s=\sum_{r=0}^{(m-1)/2} [\epsilon_1+\epsilon_2+\ldots+
\epsilon_{2r}] .\e$$
Denote by $\lambda_r=[\epsilon_1+\epsilon_2+\ldots+\epsilon_r]$. It is
easy to compute the Casimir $c_r=\lambda_r(\lambda_r+2\rho)=r(2m-r)$.
For even $m$ we then find the poles of $F(\theta)$ at
$$p_r={\zeta_r\over\lambda}={c_{r+2}-c_r\over c_2-c_0}={m-1-r\over m-1}.\e$$
In addition, there are crossing channel poles at $1-p_r=r/(m-1)$.
We conclude that the scalar $s-s$ scattering amplitude is given by,
$$S_{ss}(\theta)=\prod_{r=1}^{m-2} f_{r/(m-1)}(\theta)\cdot \wiggle
F(\theta),\e$$
where, as usual $\wiggle F(\theta)$ is a Z-factor with no poles or zeros
in the physical sheet.

For odd $m$ we find the poles at
$${\zeta_r\over \lambda}={c_{r+2}-c_r\over c_2-c_0}={m-1-r\over m-1},
\qquad {\rm for \ }r=1,3,\ldots,m-2,\e$$
along with poles from the cross channel $s \times \bar s$ at
$$1-{\bar\zeta_r\over\lambda}=1-{c_{r+2}-c_r\over c_2-c_0}={r\over m-1}
\qquad {\rm for\ } r=2,4,\ldots,m-2.\e$$
It follows that, again, the $s-s$ amplitude is
$$S_{ss}(\theta)=\prod_{r=1}^{m-2} f_{r/(m-1)} (\theta).\e$$

The soliton scattering amplitude, eq. (38) is identical to the $s-s$ scattering
amplitudes of the $D_m$ diagonal $S$ matrix, which in turn, is the same,
up to $Z$ factors, as the scattering amplitudes of the affine $D_m$ Toda
theory.
The entire set of amplitudes, which is in correspondence with the
fundamental weights of $D_m$, is given by the RSOS models based on
IRF$(D_m,\lambda_a,\lambda_a)$, where $\lambda_a$ is the $a$'th
fundamental weight, and can be found by bootstraping the $s-s$ amplitude,
or the vector amplitude, IRF$(D_m,v,v)$, which was treated in ref. \r\RSOS.
It follows that the isospin singlet part of the amplitudes is given precisely
by the $D_m$ diagonal amplitudes, which are also in correspondence with the
fundamental weights of $D_m$, and the spectrum is identical, thus, to that of
the $D_m$ diagonal theory,
$$m_a=2\sin {\pi a\over 2(n-1)},\qquad a=1,2,\ldots,n-2,\qquad
M_s=M_{\bar s}=1.\e$$

{\it $E_6$ case.} Here we consider the model IRF$(E_6,[27],[27])$,
where $[27]$ is the $27$th dimensional representation of $E_6$. Here, we
need to consider the tensor products,
$$[27]\times [27]=[\bar{27}]+[351]+[351^\prime],$$
$$[27]\times [\bar{27}]=[1]+[78]+[650],\e$$
where again, we labeled the representations by their dimension. The second
Casimirs of the representations $[1]$, $[27]$, $[78]$, $[351]$, $[351^\prime]$,
and $[650]$, are, respectively, $0$, $26/3$, $12$, $50/3$, $56/3$, and $18$.
It follows that the crossing parameters are
$$\zeta_1={\pi\over2} {8\over k+12},\qquad \zeta_2={\pi\over2}{2\over k+12},
\qquad \lambda=\bar\zeta_1={\pi\over2}{12\over k+12},$$
$$\bar \zeta_2={\pi\over2}{6\over k+12}.\e$$
The poles are thus at $\zeta_i/\lambda$, $i=1,2$, and $1-\bar\zeta_2/\lambda$,
and
$$p_1={\zeta_1\over\lambda}={2\over3},\qquad p_2={\zeta_2\over\lambda}=
{1\over 6},\qquad p_3=1-{\bar\zeta_2\over\lambda}={1\over2}.\e$$
Thus, the scalar amplitude is,
$$S_{27,27}(\theta)=f_{1/6}(\theta) f_{1/2}(\theta) f_{2/3}(\theta),\e$$
which is identical to the diagonal fundamental $E_6$ amplitude,
from which all the other
amplitudes can be found by bootstrap. It follows that the scalar part of the
$E_6$ amplitudes is given, as for the other simply laced algebras, by
the diagonal $E_6$ amplitudes, and thus, also, the mass spectrum is identical
to that of the $E_6$ diagonal amplitude.

{\it $E_7$ case}. Here we consider the model IRF$(E_7,56,56)$. The relevant
tensor product is,
$$[56]\times[56]=[1]+[133]+[1539]+[1463],\e$$
where we arranged the representations by increasing conformal dimension.
The second Casimirs of the representations, $[1]$, $[133]$, $[1539]$ and
$[1463]$, are, respectively, $0$, $18$, $28$, and $30$. Thus, the crossing
parameters are
$$\lambda=\zeta_1={\pi\over2}{18\over k+18},\qquad \zeta_2={\pi\over2}{10
\over k+18},\qquad \zeta_3={\pi\over2}{2\over k+18},\e$$
with poles at $p_2={\zeta_2\over\lambda}={5\over9}$ and
$p_3={\zeta_3\over\lambda}={1\over9}$, to which we need to add the cross
poles at $1-p_2={4\over9}$ and $1-p_3={8\over9}$, since $E_7$ is a real
group. Thus, the scalar amplitude is
$$S_{6,6}=f_{1/9}(\theta) f_{4/9}(\theta) f_{5/9}(\theta) f_{8/9}(\theta),\e$$
which is identical to the fundamental $E_7$ diagonal amplitude. Thus,
by bootstrap, all the scalar amplitudes are given by the diagonal $E_7$ ones,
along with the mass spectrum.

{\it $E_8$ case}. Here we consider the model IRF$(E_8,[248],[248])$, where
$[248]$ is the adjoint representation. We have the tensor product,
$$[248]\times[248]=[1]+[248]+[3875]+[30380]+[27000].\e$$
The second Casimirs of the representations on the r.h.s. of eq. (47) are,
respectively, $0$, $30$, $48$, $60$, $62$. We thus find the crossing
parameters,
$$\lambda=\zeta_1={\pi\over2}\cdot {30\over k+30},\qquad \zeta_2=
{\pi\over2}\cdot {18\over k+30},$$
$$\zeta_3={\pi\over2}\cdot{12\over k+30},\qquad\zeta_4={\pi\over2}\cdot{2\over
k+30}.
\e$$
It follows that the poles are at $p_2=\zeta_2/\lambda={3\over 5}$, $p_3=
\zeta_3/\lambda={2\over5}$, and $p_4=\zeta_4/\lambda={1\over15}$. These are
exactly the poles appearing in the diagonal $E_8$ amplitude
$$S_{11}(\theta)=f_{1/15}(\theta) f_{14/15}(\theta) f_{2/5}(\theta) f_{3/5}
(\theta) f_{1/3}(\theta) f_{2/3}(\theta),\e$$
with the exception of the self interaction pole at $\theta=1/3$ (and its
crossing pole at $1-\theta$). The reason for this missing pole is not
entirely clear, and might mean a breakdown in our assumption that the IRF
Boltzmann weight is given by eq. (1), or that there is really a difference
with respect to the $E_8$ diagonal theory. Note that, the mass ratios are
still given by the $E_8$ diagonal values.

This exhausts the simply laced algebras. We see that in all the cases
the assumption of the IRF Boltzmann weight eq. (1) appears to be consistent,
and that we find a sensible spectrum. In all the cases the mass spectrum is
identical to that of the corresponding affine Toda theory, while the
scalar part of the scattering matrix is identical to the corresponding
diagonal scattering theory. Note however that some problems arise for
$E_8$ because of the self interaction pole.

We turn now to the non--simply laced cases. We find essentially the same
result, that the mass spectrum is that of the classical affine Toda theory,
while the scalar $S$ matrices agree with the conjectured classical Toda
$S$ matrices
\REF\Braden{H.W. Braden, E. Corrigan, P.E. Dorey and R. Sasaki,
Nucl. Phys. B338 (1990) 689}
\REF\Deltwo{G.W. Delius,
M.T. Grisaru, S. Penati and D. Zanon, Nucl. Phys. B359 (1991)125}
\REF\Delius{G.W. Delius,
M.T. Grisaru and D. Zanon, Nucl. Phys. B382 (1992) 365}
refs. \r{\Braden,\Deltwo,\Delius}.
\REF\Destri{C. Destri, H.J. deVega and V.A. Fateev, Phys.Lett. 256B (1991) 173}
However, these $S$ matrices are known not to be consistent
at the quantum level, requiring additional particles ref.
\r{\Delius,\Deltwo,\Destri}.
It is, similarly, unclear, thus, whether the RSOS amplitudes are
consistent without additional particles.

{\it $G_2$ case}. Let us consider now the case of $G_2$. For the model
IRF$(G_2,[7],[7])$, we find the tensor product,
$$[7]\times [7]=[1]+[7]+[14]+[27].\e$$
The Casimirs of the representations, $[1]$, $[7]$, $[14]$, and $[27]$ are
respectively, $0$, $2$, $6$ and $14/3$. Thus the crossing parameters are,
$$\lambda=\zeta_1={\pi\over2} {2\over k+4},\qquad \zeta_2=\zeta_1,\qquad
\zeta_3={\pi\over2}{2\over 3(k+4)}.\e$$
We thus have a pole at $p_3={\zeta_3\over\lambda}={1\over3}$. Adding a crossing
pole at $1-p_3$ we arrive at the amplitude,
$$S_{11}(\theta)=f_{1/3}(\theta) f_{2/3}(\theta).\e$$
The pole corresponds to a bound state at $M_2=2 M_1 \cos(\pi/6)=\sqrt3 M_1$.
This is exactly the mass ratio of the $G_2$ Toda theory. The amplitude, eq.
(52), has been suggested to describe the light particle in the $G_2$
Toda theory.
(We took the cube root of the amplitudes of ref. \r{\Delius,\Deltwo}.
This does not affect the mass spectrum or the bootstrap.)

{\it $F_4$ case}. For the model IRF$(F_4,[26],[26])$ we find the tensor
product,
$$[26]\times [26]=[1]+[26]+[52]+[273]+[324].\e$$
The Casimirs of the representations on the r.h.s. are $0$, $6$, $9$, $12$
and $13$, respectively. The crossing parameters are
$$\lambda=\zeta_1={\pi\over2}{6\over k+9},\qquad \zeta_2={\pi\over2}{3\over
k+9},$$
$$\zeta_3={\pi\over2}{3\over k+9},\qquad \zeta_4={\pi\over2}{1\over k+9}.\e$$
We thus have poles at $p_2=\zeta_2/\lambda=\half$, $p_3=\zeta_3/\lambda=
\half$ and $p_4=\zeta_4/\lambda={1\over6}$. Thus the amplitude becomes,
$$S_{11}(\theta)=f_{1/6}(\theta) f_{1/2}(\theta)^2 f_{5/6}(\theta).\e$$
When comparing with affine Toda results, we miss a self interaction pole
at $\theta=1/3,2/3$. The Toda amplitude is
$$S_{11}(\theta)=f_{1/6}(\theta) f_{1/3}(\theta) f_{1/2}(\theta)^2 f_{2/3}
(\theta) f_{5/6}(\theta).\e$$
As in the $E_8$ case, the meaning of this missing pole is unclear. It might
either signify a breakdown of the anzats for the Boltzmann weight, eq. (1),
or, a small difference with respect to the conjectured Toda amplitude.
We cannot just add the self interaction pole, since this will lead to problems
in the bootstrap of the full amplitude (see the following).

{\it $B_n$ case}. Consider the model IRF$(B_n,s,s)$, where $s$ stands for the
spinor representation. Denote the simple roots of $B_n$ by $\alpha_r=
\epsilon_r-\epsilon_{r+1}$, $r=1,2,\ldots,n-1$, $\alpha_n=\epsilon_n$,
where the $\epsilon_i$ form an orthonormal set of vectors. Then,
the spinor highest weight is,
$$s=[{\epsilon_1+\epsilon_2+\ldots+\epsilon_n\over2}],\e$$
and we have the tensor product,
$$s\times s=\sum_{m=0}^n [\epsilon_1+\epsilon_2+\ldots+\epsilon_m].\e$$
For the Casimirs we find,
$$C_{[\epsilon_1+\epsilon_2+\ldots+\epsilon_m]}=\half m(2n+1-m).\e$$
Thus the crossing parameters are
$$\lambda=\zeta_1,\qquad
\zeta_m={\pi\over 2} \cdot {n-m+1 \over 2n-1+k},\qquad m=1,2,\ldots,n.\e$$
The poles are thus located at
$$p_m={\zeta_m\over\lambda}={n-m+1\over n}\qquad m=2,3,\ldots n.\e$$
Note, however that to these poles we need to add the cross channel poles
located
at $1-p_m$, which so happen to coincide exactly with the original set of
poles $p_m$. Thus the full scalar amplitude becomes,
$$S_{ss}(\theta)=\prod_{m=1}^{n-1} f_{m/n}(\theta)^2.\e$$
Note the appearance of the perplexing double poles as a result of the
coincidence of the direct and cross channel poles. In fact, exactly the same
amplitude, $S_{ss}$ appears in the $B_n$ affine Toda theory.  The perplexing
double poles create a problem there, as well. One way to solve the problem
was suggested \r{\Delius,\Deltwo,\Destri}\  is to add a fermion to the theory.
The double poles then
arise naturally from fermion loops. Presumably, the same kind of solution is
needed here, namely, we are missing a fermion particle in the theory. More
study
of this is required.

Since all the other amplitude appear as a bootstrap of the fundamental
$S_{ss}$, we find an agreement between all the amplitudes and those of
affine Toda theory. The model IRF$(B_n,\lambda_m,\lambda_m)$, where
$\lambda_m$ is the $m$th fundamental weight, corresponds to the
scattering of the $m$th Toda particle, or the $\lambda_m$ kink, whose
mass is given by
$$M_m=2\sin({\pi m\over 2n}), \qquad m=1,2,\ldots,n-1,\qquad M_s=1.\e$$
The vector amplitude was considered directly in ref. \r\RSOS\ and, again, a
perfect
agreement with affine Toda theory is found.

{\it $C_n$ case.} Here we consider the model
IRF$(C_m,[\lambda_1],[\lambda_1])$,
where $\lambda_m$ is the $m$th fundamental weight (`vector')
\foot{This model was considered also in ref. \r\RSOS. An error was made
there in
the comparison with the $C_n$ model of ref.
\REF\Jimbo{M. Jimbo, T. Miwa and M. Okado, Lett.Math.Phys. 14 (1987) 123}
\r\Jimbo. However, as we see here,
the agreement with affine Toda still holds perfectly.}.
The relevant tensor product is
$$[\lambda_1]^2=[0]+[\lambda_2]+[2\lambda_1],\e$$
where we denoted the representations by their highest weight.
The conformal dimensions of the fields are,
$$\Delta_{[2\lambda_1]}={n+1\over k+n+1},\qquad \Delta_{[\lambda_2]}=
{n\over k+n+1},\e$$
and the crossing parameters become,
$$\lambda=\zeta_1={\pi\Delta_{[\lambda_2]}\over2}={\pi n\over 2(k+n+1)},\qquad
\zeta_1={\pi\over2}(\Delta_{[2\lambda_1]}-\Delta_{[\lambda_2]})=
{\pi\over2(k+n+1)}.\e$$

Thus, there is a unique pole at $p_1=\zeta_1/\lambda={1\over n}$, along with
the cross channel pole at $1-p_1={n-1\over n}$. The scalar amplitude thus
becomes,
$$S_{11}(\theta)=f_{1/n}(\theta) f_{1-1/n}(\theta).\e$$
This is exactly the fundamental amplitude of the $C_n$ Toda theory
from which all
the other amplitudes can be obtained by bootstrap. The mass spectrum is
also the same as the Toda one, with the identification of the $m$ particle
with the $\lambda_m$ kink,
$$M_m=2\sin({\pi m\over 2n}),\qquad m=1,2,\ldots,n.\e$$
Again, a double pole problem arises in some amplitudes which can be solved,
in the Toda case, by adding additional two fermions ref. \r{\Delius,\Deltwo,
\Destri}.
As the same problem arises here, presumably, the analogous particles need
to be added.

This concludes all the algebras. We find, in all cases, an agreement with
the spectrum of the corresponding affine Toda theory. The number of
kinks is equal to the rank of the algebras. Their masses are given by the
eigenvector with maximal eigenvalue (Perron Frobenius) vector of the
Cartan matrix. The kinks are in one to one correspondence with the
simple weights of the algebra. The admissibility condition with respect
to the $\lambda$ kink, where $\lambda$ is a fundamental weight, is given by
fusion with respect to $\lambda$, i.e., $a\sim b$ iff $a\lambda=b+\ldots$,
where $a$ and $b$ are any weights. The scattering amplitude of two kinks
contains a factor which is identical to the affine Toda scattering amplitude,
which is responsible for all the poles in the physical sheet.

We wish to study now the bootstrap of the full RSOS amplitudes. Consistency
requires that under bootstrap the set of amplitudes will be closed. We
wish to judge if this is the case.
Consider then the kink scattering process,
$$K_{\alpha,\beta}+K_{\beta,\gamma}\rarrow K_{\alpha,\delta}+K_{\delta,\gamma}
,\e$$
where $K_{\alpha,\beta}$ denotes the $K$ kink interpolating between the
$\alpha$ and $\beta$ vacua. The scattering amplitude for this process
is denoted by $S^{\beta,\gamma}_{\alpha,\delta} (\theta)$, where
$i\pi \theta$ is the relative rapidity. The poles in $S$ on the
physical sheet, $0<\theta<1$ correspond to bound state kinks. The coupling
to the bound state, $g_{\alpha\beta\gamma}$, is given by,
$$g_{\alpha\beta\gamma} g_{\alpha\delta\gamma}=\pi
\Res_{\theta=\theta_0} S^{\beta\gamma}_{\alpha,\delta} (\theta),\e$$
where $0<\theta_0<1$ is the location of the pole.

Denote the bound states kink by $B$. Then the scattering matrix of $B$ and the
original kink, $K$, which is denoted by $H^{\gamma\delta}_{\alpha\eta}$, is
given by the bootstrap equation, ref. \r\ZamZam
$$ g_{\eta\epsilon\delta} H^{\gamma\delta}_{\alpha\eta}=
\sum_\mu g_{\alpha\mu\gamma} S^{\gamma\delta}_{\mu\epsilon}(\theta+\theta_p/2)
S^{\mu\epsilon} _{\alpha\eta}(\theta-\theta_p/2).\e$$

Recall the anzats for the IRF scattering theory, eq. (6),
$$S^{ab}_{cd}(\theta)=F(\theta) \left({S_{a,0} S_{d,0} \over S_{b,0} S_{c,0}}
\right)
^{-\theta/2} \sum_{\alpha=1}^n f_\alpha(\theta) P^\alpha\pmatrix{a&b\cr c&d
\cr},\e$$
where the functions $f^\alpha(\theta)$ are given by eq. (2).

Consider now the poles of $F(\theta)$ which would lead to poles in the
full $S$ matrix. These are located at the points $0<\theta_l<1$, where
$\theta_l=\zeta_l/\lambda$,  along with possible poles from the cross
channel. Now, from the expression for $f^a(\theta)$, eq. (2), we find the
important property, $f^a(\theta_l)=0$, for $a=l,l+1,\ldots,n-1$. Thus the
coupling to the bound states in the channel with the pole $\theta_l$
becomes,
$${1\over\pi} g_{\alpha\beta\gamma}^2=\left[\Res_{\theta=\theta_l}
F(\theta)\right]
\left[{S_{\beta,0}^2 \over S_{\alpha,0}S_{\gamma,0}}\right]^{-\theta_l/2}
\sum_{a=0}^{l-1}
f^a(\lambda\theta_l) P^a\pmatrix{\beta&\gamma\cr\alpha&\beta\cr}.\e$$
Note, importantly, that only projection operators, $P^a$,  with $a<l$
contribute to the coupling. However, $P^a$ corresponds in RCFT to the
projection of the braiding matrix on the intermediate $t$ channel involving
the field $\phi_a$. Thus, $P^a\pmatrix{\beta&\gamma\cr\alpha&\beta\cr}
\neq0$ only if $\alpha\times \phi_a=\gamma+\ldots$, where the product is in
the fusion rules sense. It follows that the bound state $B$ obeys the
admissibility condition, $B_{\alpha,\gamma}\neq0$ only if $\alpha t=\gamma+
\ldots$ where $t=\phi_0,\phi_1,\ldots,\phi_{l-1}$.

Thus, the admissibility condition for the IRF model obtained from the
bootstrap equation, for the field $B$, is given by fusion with respect to the
field, $\phi=\phi_0+\phi_1+\ldots+\phi_{l-1}$, where these are the fields
appearing in the OPE, $x^2=\sum_a \phi_a$, arranged according to increasing
conformal dimensions. For the examples at hands it can be observed that
the admissibility condition is equivalent to fusion with respect to the
field $\phi_{l-1}$, since $\alpha\phi_r=\gamma+\ldots$ with $r<l-1$,
implies that $\alpha\phi_{l-1}=\gamma+\ldots$. Thus, we conclude that the
bound states $B$,  $S$ matrix, is described by the model IRF$({\cal O},
\phi_{l-1},\phi_{l-1})$, and, importantly, is also a fusion IRF model.
In summary, the bound states of the kink $x$ with itself at the pole
$\zeta_l/\lambda$ is described by the field $\phi_{l-1}$ appearing in
the operator product of $x$ with itself.

Let us turn now to examples.

(1) $SU(n+1)$. Consider the scattering process of $x=\Lambda_m$. We found
above the unique pole at
$$\theta_m={\zeta_0\over\lambda}={2m\over n+1},\e$$
where $\zeta_0$ corresponds to the tensor product $x^2=\Lambda_{2m}+\ldots$.
Since $f_1(\theta_m)=f_2(\theta_m)=\ldots=0$, as explained above, it
follows that the bound state $S$ matrix corresponds to the model
IRF$(A_n,\Lambda_{2m},\Lambda_{2m})$. This is in agreement with the Toda
picture, since, as explained above, the $\Lambda_r$ representation is in
correspondence with the $r$ Toda particle, and the process in the
Toda theory is 	$m+m\rarrow 2m$. Thus, the bound state representation, for
consistency, has to be $\Lambda_{2m}$. This, agrees with the direct
calculation using the bootstarp equation.

(2) $D_m$. For even $m$ we have $s=\bar s$ and the scattering of two
spinor kinks is real. We have the OPE, eq. (31),
$$s^2=\sum_{r=0}^{m/2} [\epsilon_1+\epsilon_2+\ldots+\epsilon_{2r}],\e$$
The poles are located (see above) at
$$p_r={m-1-2r\over m-1},\qquad r=1,2,\ldots,{m\over2}-1.\e$$
The $p_r$ pole gives the representation $x=[\epsilon_1+\epsilon_2+\ldots+
\epsilon_{2r}]=\Lambda_{2r}$ as $f^{a+1}(p_r)=0$ for $a>r$. Thus, the bound
states corresponds to the RSOS theory IRF$(D_m,x,x)$. The mass of this
particle is,
$${M_B\over M_s}=2\cos ({\pi p_r\over2})=2\sin({\pi r\over m-1}),\e$$
which is exactly the mass of the $\Lambda_{2r}$ particle in the Toda theory.
Thus, we get the correct representation according to the correspondence
with Toda theory.
For $D_m$, odd $m$, the calculation is similar. Again, we find the correct
representations in the bootstrap.

It is similarly straight forward to see that for the Lie algebras $B_n$,
$C_n$, $E_6$, $E_7$ and $G_2$, the fundamental amplitudes bootstarp correctly,
and indeed the correct representations arise. We encounter problems, however,
for the algebras $F_4$ and  $E_8$. We already saw that, in these
cases, we miss a
self interaction pole at $\theta=2/3$.
Furthermore, the other poles do not
bootstrap correctly, i.e. wrong representations arise. We conclude that
some modification of the anzats eq. (1) is required in these cases. This is
the most logical conclusion, since the RSOS soliton picture should persist.
Curiously, all the other poles are found to be in the correct place, so even
for these problematic cases the anzats, eq. (1), seems `almost' right.

The quantum field theory described by the algebra $G$ RSOS scattering theory,
with the simple weights for the admissibility condition, was conjectured
to be the coset theory $G_k\times G_1\over G_{k+1}$  \r{\Vega,\RSOS},
perturbed by the field $\Phi^{0,0}_{\rm ad}$. Evidence for this is that
these perturbed cosets have the correct spins for the integrals of
motions, which are the exponents of the Lie algebra modulo the Coxeter
number. Further in ref.
\REF\Ahn{C. Ahn, D. Bernard and A. LeClair, Nucl. Phys. B346 (1990) 409}
\r\Ahn, these cosets were mapped directly to Toda theories, and it was
argued, for $A_n$, that the mass spectrum is the same as the corresponding
Toda theory.
This is additional evidence for this identification.

Finally, the anzats eq. (1) seems to be consistent in almost all the
cases we studied. We find the correct poles and the correct bootstarp
structure. This is a highly constrained check which lends considerable
support for this anzats.

We hope that this work will be of benefit in the understanding of solvable
lattice model and the related RSOS scattering theories.

\refout
\bye